# Is gravity inherent to relativistic many-particle systems?

Niclas Thisell
email: niclas@thermometric.se
www: http://www.thermometric.com/private/niclas/phys.html
revision: February 23, 2000

## *Abstract*

This paper discusses the somewhat unintuitive conjecture that many Lorentz-invariant many-particle models can be reinterpreted to satisfy the gtr field equations. It is shown that a careful remapping of coordinates yields a non-trivial Riemannian manifold. Furthermore an energy-momentum tensor is outlined and it is argued that it may converge to its classical counterpart in the macroscopic limit. These ideas could possibly be used to partially relieve us of some the resilient problems of adding spacetime curvature to modern QM theories.

## (1) Motivation

The basic idea is, simply put; to reinterpret the coordinates of particles for a model defined in a flat many-particle Hilbert-space. Or more precisely, we start with a *N*-particle model defined in a space having 4*N* dimensions. This space is then mapped into another 4*N*-dimensional space. We proceed to define parallel coordinate systems, allowing us to define a comma derivative. Using similar techniques, we can also propose a way to transform tensors from the original flat space. The comma derivative effectively defines 4-dimensional slices from the 4*N*-dimensional space. And the model is shown to be valid in these slices. Now, looking closely at one of these subspaces, the metric shows us that it is curved. Which means that the model looks just as valid when looking at it from a curved space-time.

While all this may sound trivial, it would actually imply that our space-time curvature is a matter of interpretation.

## (2) Introduction

Let's start with a relativistic model *M* using the coordinate system $x_1...x_N$ in Minkowski-space, where each $x_I$ is a 4-coordinate (the 'many-times formalism', apparently already considered by Dirac [1]). The exact requirements for *M* cannot be stated until we have developed the necessary terminology and notation. Now, let us temporarily revert back to a space with common time and look at the coordinate system $\tilde{x}_1...\tilde{x}_N$ and *t*, where each $\tilde{x}_I$ is a 3-coordinate and *t* is time. There is an infinite number of ways to select this subvolume, but, for now, let us simply set $x_{Io} = t$ for all particles and $\tilde{x}_{Ii} = x_{Ii}$ (where *i* spans from 1 to 3)

Then define

(2.1) $y^I = y^I(t, \tilde{x}_1...\tilde{x}_N)$

where $y^1...y^N$ should be interpreted as the space-time coordinates of all particles. Or more precisely, $y^{Ij}$ is the 4-coordinate of particle *I*.

Let us examine what happens when differentiating this function.

(2.2) $\partial y^{Ii} = \dfrac{\partial y^{Ii}}{\partial t} \partial t + \sum_K \dfrac{\partial y^{Ii}}{\partial \tilde{x}_{Kj}} \partial \tilde{x}_{Kj}$

Note the convention of implicitly summing over lower case letters, while not automatically summing over particle indices, which will be assigned capital letters.

Now, let's ponder how differentiation in *y* should be calculated. How do we know how much the $\tilde{x}$-coordinates move when we nudge a single y-coordinate? The problem is that the *y*-space is a lot bigger than its counterpart in $\tilde{x}$. An analogy in three dimensions would be differentiating $q$ and $j$ of a sphere with respect to the z-axis. We can't simply increase the z-coordinate to see what happens – at most points we will leave the surface. So we must also nudge the x- or y-coordinate to stay in the sphere. But there is no unique way of doing that. So let's try to handle all possible ways of doing it for a while.

We can start by defining

(2.4) $D^{IiJ0} = \dfrac{\partial y^{Ii}}{N \partial t}, D^{IiJj} = \dfrac{\partial y^{Ii}}{\partial \tilde{x}_{Jj}}$

We can then rewrite (2.2)

(2.5) $\partial y^{Ii} = \sum_K D^{IiKj} \partial x_{Kj}$

Now, the equivalent of the falling-off-the-sphere-problem could be stated as not being able to move a single particle without also having to move some other particles. To treat this thoroughly we should define yet another coordinate system.

(2.6) $x_{Jj} = x_{Jj}(\underset{1}{z}...\underset{N}{z}, \underset{1}{w}...\underset{N}{w})$

where every $\underset{I}{z}$ and $\underset{I}{w}$ is a 4-coordinate. (The somewhat oddball placement of the indices will pay off later when we need to easily spot the difference between various coordinate system indices). So again we define

(2.7) $B_{Kk}^{Jj} = \dfrac{\partial x_{Kk}}{\partial \underset{Jj}{z}}$

so that

(2.8) $\partial x_{Kk} = B_{Kk}^{Jj} \partial z_{Jj}$

Now, when designing this function we should try to make sure that we could find an inverse for a single particle. Let's implicitly define, while also requiring that it can always be calculated

(2.9) $B_{Jj}^{Jk} B_{Ji}^{Jj} = d_i^k$

(The right-hand side is the Kronecker delta with our z-coordinate-system index placement).
This means that we can write

(2.10) $\partial_{(z)}_K x_{Jj} = B_{Jj}^{Kk} B_{Kk}^{Ki} \partial x_{Ki}$

Note that $x_J$ is certainly not just a function of $z_K$ - the $\partial_{(z)_K} x_J$-notation is just meant to emphasize the fact that we're differentiating by varying that single $z_K$-coordinate. (2.5) becomes

(2.11) $\partial_{(z)_J} y^{Ii} = \sum_K D^{IiKk} B_{Kk}^{Jl} B_{Jl}^{Jj} \partial x_{Jj}$

That means that we have a differentiation defined in terms of coordinates of a single particle. We can make it even shorter by defining

(2.12) $C_{Kk}^{JJj} = B_{Kk}^{Jl} B_{Jl}^{Jj}$

The extra J-index serves to make a distinction from its inverse, even though we won't actually define the inverse here. Anyway, now let us define

(2.13) $\overline{D}^{IiJ,j} = \sum_K D^{IiKk} C_{Kk}^{JJj}$

enabling us to rewrite (2.11) again

(2.14) $\partial_{(z)_J} y^{Ii} = \overline{D}^{IiJ,j} \partial x_{Jj}$

Notice the introduction of the comma before the 'j'-index. This could be interpreted as a new type of differentiation, roughly meaning the derivative along an axis of a particle, while also moving the other particles as specified by the B-matrices. Anyway, this leaves us in a position to define an inverse

(2.15) $\overline{D}_{IiI,j} \overline{D}^{IjI,k} = d_i^k$

and get

(2.16) $\partial_{(z)_I} x_{Ij} = \overline{D}_{IjI,i} \partial y^{Ii}$

Note that, while doing these comma-derivatives in y-space, the coordinates of the other particles move, much like in x-space. In a sense, the B-matrices define the parallel coordinate system of a particle with respect to the coordinate system of another particle. And the new differentiation operator defines the derivate in this related coordinate system.

So, what do we do now? We could start trying to generalize the differentiation concept to not only handle our coordinates. How would we define the derivative of a scalar? To make the notation less cumbersome we should try to focus on a single particle I. The logical extension of the previous definitions would be

(2.17) $\boldsymbol{f}^{,l} = \sum_K \frac{\partial \boldsymbol{f}}{\partial x_{Kk}} \frac{\partial_{(z)_I} x_{Kk}}{\partial x_{Il}} = \sum_K \frac{\partial \boldsymbol{f}}{\partial x_{Kk}} C_{Kk}^{Il}$

and

(2.18) $\boldsymbol{f}_{,l} = \overline{D}_{ImI,l} \boldsymbol{f}^{,m}$

Note that these comma-derivatives may and probably will differ for different particles, even though we're looking at the same scalar field. (We could have used a more exact notation, like $\boldsymbol{f}_{,(Ik)}$, but that quickly clutters up our equations. So let's not).
Now define

(2.19) $\hat{\boldsymbol{f}}(z_1,...z_N) = \boldsymbol{f}(x_1(z_1...z_N),...x_N(z_1...z_N))$

and get

(2.20) $$\frac{\partial \hat{f}}{\partial z_{Ii}} = \sum_K \frac{\partial f}{\partial x_{Kk}} \frac{\partial x_{Kk}}{\partial z_{Ii}} = \sum_K \frac{\partial f}{\partial x_{Kk}} \frac{\partial x_{Kk}}{\partial z_{Ii}} = \sum_K \frac{\partial f}{\partial x_{Kk}} B_{Kk}^{Ii}$$

Then use (2.10) to get

(2.21) $$\frac{\partial \hat{f}}{\partial z_{Ii}} = \left( \sum_K \frac{\partial f}{\partial x_{Kk}} B_{Kk}^{Im} B_{Im}^{Il} \right) B_{Il}^{Ii} = f^{,l} B_{Il}^{Ii}$$

This leads us to define

(2.22) $$\overset{,l}{f} = f^{,i} B_{Ii}^{Il}$$

Now, since we're never actually using the absolute z-coordinates (only the interparticle coordinate system matrices it implicitly defines), we can actually set

(2.23) $$B_{Jj}^{Jk} = d_j^k$$

and (2.9) will give us

(2.24) $$B_{Ji}^{Jj} = d_i^j$$

That makes $\overset{,l}{f}$ and $f^{,l}$ virtually identical and equally well behaved. And since $\overset{,l}{f}$ quite plainly was the differential with respect to a single coordinate, higher order differentials should be reasonably nice. For instance the order in which things are differentiated won't matter.

**(2.1) Differentiation operator**

Ok, we would also like a pure differentiation operator, like $\partial/\partial x_{Jj}$. Let us do this using similar techniques, while possibly clarifying the methods used. We should start by defining

(2.1.1) $$x_{Ii} = w'_{Ii} + \sum_K z'_{Ki}$$

This will give us an identity $C'^{IIi}_{Kk}$ and we can define

(2.1.2) $$\partial^{IJi} = \frac{\partial(z'_I) x_{Jj}}{\partial x_{Ii}} \frac{\partial}{\partial x_{Jj}}$$

and get (exactly what we were after)

(2.1.3) $$\partial^{IJi} = \frac{\partial}{\partial x_{Ji}}$$

Now, using the exact same techniques as before, we can define

(2.1.4) $$\overline{D}'^{Iijj} = \sum_K D^{IiKk} C'^{Jj}_{Kk} = \sum_K D^{IiKj}$$

This tells us how the *y*-coordinates move when nudging the *x*-coordinates using this totally parallel coordinate system. Like this

(2.1.5) $$\partial(z'_J) y^{Ii} = \overline{D}'^{IiJj} \partial x_{Jj}$$

This leads us to define a y-space counterpart of $\partial^{IJi}$

(2.1.8) $\partial_{IJi} = \dfrac{\partial(z'_I)x_{Jj}}{\partial y^{Ii}}\dfrac{\partial}{\partial x_{Jj}} = \dfrac{\partial(z'_I)x_{Ia}}{\partial y^{Ii}}\dfrac{\partial(z'_I)x_{Jj}}{\partial x_{Ia}}\dfrac{\partial}{\partial x_{Jj}}$

and it becomes

(2.1.9) $\partial_{IJi} = \overline{D}'_{IaIi}\partial^{IJa}$

**(2.2) Time**

We introduced the coordinates $\tilde{x}_1...\tilde{x}_N$ and $t$ to get a better feel what parallel coordinates mean. But now is the time to go back to pure 4-coordinates in flat space as well. It shouldn't pose a big problem, though, since we've been using the $x_I$-coordinates throughout, while pretending that they have common time. Some definitions need to be tweaked (for the better). For instance (2.1) becomes

(2.2.1) $y^I = y^I(x_1...x_N)$

and (2.4) becomes

(2.2.2) $D^{IiJj} = \dfrac{\partial y^{Ii}}{\partial x_{Jj}}$

But not a lot changes. Except for the fact that we have no time at all. (The reader may still think of 3N-dimensional subvolumes as having equal 'time' in some sense. It should also be pointed out that this approach is more general – i.e. we can still design the functions so that time is common).

## *(3) Tensors*

Ok, now we could try to define a couple of tensor transformation rules. But before writing down the actual formulae it should be pointed out that tensor indices in x are upside-down. This is confusing at first, and would not have been a problem had papers had more than 2 dimensions. But they don't, so contravariant indices in x are written subscript and vice versa.

Anyway, the structure of (2.1.9) leads us to define covariant tensors like this

(3.1) $A_i = \overline{D}'_{IjIi}\breve{A}^j$

And conversely, the contravariant x-tensor $\breve{A}_j$ must become

(3.2) $A^i = \overline{D}'^{IiIj}\breve{A}_j$

Note again that these tensors very much depend on which particle we are studying. Now, let's practice by calculating the metric as seen by a single particle. We know that the x-space has a simple Minkowski-metric.

$$\breve{g}^{ij} = \breve{h}^{ij} = \begin{bmatrix} -1 & 0 & 0 & 0 \\ 0 & 1 & 0 & 0 \\ 0 & 0 & 1 & 0 \\ 0 & 0 & 0 & 1 \end{bmatrix}^{ij}$$

That gives us

(3.3) $g_{ij} = \overline{D}'_{IkIi}\overline{D}'_{IlIj}\breve{h}^{kl}$

and

(3.4) $g^{ij} = \overline{D}'^{IiIk}\overline{D}'^{IjIl}\breve{h}_{kl}$

We now have all the tools we need to calculate Christoffel symbols, covariant derivatives, Riemann curvature tensors, Ricci tensors and the Einstein tensor. Which is cool.

Before we start actually calculating anything, we could briefly consider a single particle system. Then we will have

$(N=1) \to \left(\overline{D}^{IiI,j} = \overline{D}'^{IiIj}\right)$

and get no curvature. This clearly makes sense since y->x will be a 4space->4space mapping.

### (3.1) Interactions

Let us define

(3.1.1) $\nabla_{IKa} = x_{Ia} - x_{Ka}$

This 4-component function is roughly the separation of the particles. The coordinates where $\nabla_{IK} = 0$ define points of interaction, i.e. where particle *I* 'touches' particle *K*. It would be neat to have a function whose integral over such a point returns 1. That doesn't sound very hard. Define

(3.1.2) $\boldsymbol{d}^K = \boldsymbol{d}^{(4)}(\nabla_{IK})$

where $\boldsymbol{d}$ is the Dirac delta distribution. Let's try it out

$$\int_V \boldsymbol{d}^K \frac{1}{\partial^{I0}...\partial^{I3}} = \int_V \boldsymbol{d}(x_{I0} - x_{K0})...\boldsymbol{d}(x_{I3} - x_{K3}) \frac{1}{\partial^{I0}...\partial^{I3}} = \begin{cases} 1 \text{ if } x_K \in V \\ 0 \text{ if } x_K \notin V \end{cases}$$

(The $1/\partial^{II}$ looks kind of wacky, but keep in mind that $\partial^{II}$ is defined as $\partial/\partial x_I$.) But how does this translate to *y*-space? We would obviously like the interactions at the 'same place' in *y*-space. We *could* require

(3.1.4) $(x_K = x_I) \rightarrow (y^K = y^I)$

This may be comforting, but we really don't have any use for it. Be that as it may – here, we will pretend that $\nabla_{IK}$ is a tensor and transform it to get a *y*-space counterpart $\nabla^{IK}$. Obviously, $\nabla^{IK} = 0$ if $\nabla_{IK} = 0$. This allows us to define

(3.1.5) $\boldsymbol{d}_K = \boldsymbol{d}^{(4)}(\nabla^{IK})$

Again, let us try to integrate

$$\int_V \boldsymbol{d}_K \frac{1}{\partial_{I0}...\partial_{I3}} = \int_V \boldsymbol{d}^{(4)}(\overline{D}'^{IiIj}\nabla_{IKj}) |\overline{D}'^{IaIb}| \frac{1}{\partial^{I0}...\partial^{I3}} = \begin{cases} 1 \text{ if } x_K \in V \\ 0 \text{ if } x_K \notin V \end{cases}$$

## *(4) The Model*

Now, you may recall that we never got around to defining the exact requirements of the flat relativistic quantum mechanics model *M*. Well, now is the time to try:

$\partial^{IJ}$ **must be the only operator on coordinates.**

The point being, of course, that this operator becomes a tensor in both flat *x*-space and curved *y*-space. And we should add that anything that translates nicely to *y*-space may be used, e.g. tensors and the $\boldsymbol{d}^J$-distribution. The exact motivation will not become apparent until the local inertial frame has been examined, which we will do in the next section.

Unfortunately the given requirement is pretty vague, so some further explanation is needed. First of all, operators that are derived from $\partial^{IJ}$ are also allowed, as long as they don't cause trouble. For instance, integration may be allowed, as long as it can be differentiated to be expressed using $\partial^{IJ}$. Of course, it's cause trouble with integrals as well. Integrals all over space may not be unambiguously defined here. Neither is integrations from a 'time' to another 'time'.

Technically, Lagrangian densities implicitly need integration, but since they can almost always be rewritten on pure differential form, they should work (as long as this differential form can be expressed using $\partial^{IJ}$).

The requirement is admittedly still vague. The only failsafe method is to try to apply the transformations suggested here to see if it works out (this may also require invention of spinor-transformations etc).

### (4.1) Local inertial frame

Now let us ponder the proposition suggested by Einstein [2], that we can always find a local inertial frame. Or more precisely, for every point, there exists a local coordinate system such that

(4.1.1) $\hat{g}_{ij}(p) = \boldsymbol{h}_{ij}$

and

(4.1.2) $\hat{g}_{ij,k}(p) = 0$

Looking at (3.3), this local coordinate system doesn't look all that hard to find. And it is in fact relatively easy to find an explicit Taylor expansion $\hat{y}^{Ii} = \hat{y}^{Ii}(y^I)$ around a point that satisfies (4.1.1) and (4.1.2). Using this coordinate system all tensors have been transformed back to look exactly their *x*-space counterparts (albeit upside-down). Clearly, the model *M*, using these tensors, must be satisfied in $\hat{y}$ as well. Having done that, the model *M* will locally feel and smell exactly the same in flat x-space as it does in curved $\hat{y}$-space.

The scientists in $\hat{y}$ will conclude that their model *M* is only almost good – only, real world objects seem to fall towards each other on a larger scale. And they don't in their model *M*. (Or so they thought).

## *(5) Energy-momentum*

Let us start by focusing our attention to a 4-dimensional slice of spacetime. We can define a 4-coordinate *Y* by integrating the following (it doesn't matter 'from where')

(5.1) $Y^a_{,b} = d^a_b$

Now recall that $\nabla_{IKa}$ defined points of interaction. We would like particles not to appear or disappear spontaneously, so it'd be nice if these nodes described paths in *Y*.

We pretended that $\nabla_{IKa}$ was a tensor to get $\nabla^{IKa}$. In fact, where $\nabla_{IK} = 0$, $\nabla^{IKa}$ will truly be a tensor :-). And so will $\nabla^{IKa}_{;b}$ since it will be equal to $\nabla^{IKa}_{,b}$. We can be sure that there will be a non-zero tensor $U^i$ along these nodes such that

(5.2) $(\nabla^{IK} = 0) \to (U^{IKb} \nabla^{IKa}_{;b} = 0)$

If there is a path, the 'rank' of $\nabla^{IKa}_{;b}$ will be at most 3. Unfortunately, it may also be lower, thus allowing more tensors than we had planned. For instance, if $C^{IIj}_{Kk} = d^j_k$ at an interaction (i.e. that the coordinate-systems are totally parallel), the rank will be 0. If this is the case, or if this *should* be the case, (5.2) needs to be refined.

But let us pretend that $U^i$ is decided except for magnitude. We can make it a nice 4-velocity by adding

(5.3) $(\nabla^{IK} = 0) \to (U^{IKa} U^{IKb} g_{Iab} = -1)$

Ok, now we know (or at least we're pretending that we know) the 4-velocities of the particles. Then we can define

(5.4) $P^{IKa} = m_K U^{IKa}$

where $m_K$ is the mass of particle *K*, to be the 4-momentum of particle *K*. Unfortunately, this definition doesn't work very well with mass-less particles. We could either try to ignore them and hope that they don't contribute to the energy-momentum tensor. Or we could admit that we don't know how they should be handled. Then define

(5.5) $N^{IKa} = s_{IK} U^{IKa}$

to be the particle-flow of particle *K*. $s_{IK}$ is left undefined for now, but should be thought of as a spike at interactions, i.e. it's 0 where $\nabla_{IK} \neq 0$ and pretty high where $\nabla_{IK} = 0$.

We can define

(5.6) $T^{IKab} = N^{IKa} P^{IKb}$

and then define the energy-momentum tensor for particle *I*

(5.7) $T^{Iab} = \sum_{K \neq I} N^{IKa} P^{IKb}$

And finally we set

(5.8) $G_{Iab} = T_{Iab}$

But we must also make sure that $T^{Iab}_{;b} = 0$. So let's have a look what it looks like

(5.9) $T^{IKab}_{;b} = m_K (s_{IK,b} U^{IKa} U^{IKb} + s_{IK} U^{IKa}_{;b} U^{IKb} + s_{IK} U^{IKa} U^{IKb}_{;b})$

The sum of those three terms must be set to 0. Let us look at one at a time. $s_{IK,b} U^{IKb}$ roughly means the rate of change of particle-density along a path. While we haven't defined $s_{IK}$ yet, it sounds reasonable that this change should be set to zero once we do. $U^{IKa}_{;b} U^{IKb}$ is roughly the covariant change of velocity along a path.

Setting this to zero means that the particles will follow geodesics (for a particular slice). And $U^{IKb}_{;b}$ is the divergence of the velocity. We haven't really thought about what $U^{IKa}$ should be, or even represent, outside the path. But it seems setting the divergence to zero should cause no trouble.

So, the first term should definitely be set to 0, thus leaving the sum of the remaining two terms 0. If we set all three to zero, we get nice geodesics. There is a glimmer of hope that this latter choice may give us an energy-momentum that observers will agree upon in the macroscopic limit. To decide if this is true, however, we need to examine what it feels like to be an observer here. Things easily get very hand-wavy right about here when trying to argue how this might work. We should look at a path through *y*-space and try to figure out what they will think their energy-momentum is. And if we have disagreement, one could perhaps try to avoid defeat by making a prediction. This is, however, beyond the scope of this paper (and not known if it is true).

Ok, but how should $s_{IK}$ be defined? We would like to get 1 when integrating it over a 3-volume through which particle *K* passes. Unfortunately, elegant definitions seem to have a tendency to require 0 times infinity to be equal to 1 (i.e. they are effectively not well-defined). So this task is left as an exercise for the reader.

### (5.1) Solving the equations

Finally, can these equations be solved? You may recall that we had

$$x_{Jj} = x_{Jj}(\underbrace{z...z}_{1 \quad N}, \underbrace{w...w}_{1 \quad N})$$

Now, those $w_I$ are only needed when constructing flat coordinate systems, as demonstrated in (2.1.1). And they are mostly pointless otherwise. So it is conjectured that

(5.1.1) $x_{Jj} = x_{Jj}(\underbrace{z...z}_{1 \quad N})$

We also had

(5.1.2) $y^I = y^I(x_1...x_N)$

This leaves us with *(4+4)N* free variables. And we get *10N* equations from the Einstein field equation (4 of which are 'wasted' satisfying zero divergence). So unless we're lucky, it's not enough. It is unknown if they can be solved.

## *(6) Summary*

What does it mean? An intuitive picture is thinking of it as a magnifying glass of sorts. When looking through this magnifying glass, it looks like particles gravitate towards each other - even though we haven't changed the equations of motion like the Hamiltonian or Lagrangian! And since we made sure we didn't violate any of the gtr postulates, these magnified frames should be just as valid as the untransformed flat vanilla-frame. The Hamiltonians and Lagrangians and what have you should be locally true in these transformed coordinates (using our handcrafted derivatives etc).

But why should we use this magnifying glass at all? Well, why not?

As far as physics goes it may suffice to consider the suggestion that the metric may be a function of all coordinates

(6.1) $g_{Iab} = g_{Iab}(y^1...y^N)$

Having said all that, it should be pointed out that a lot of questions remain on these issues. Can the equations be solved? Can we actually formulate any interesting models using the $\partial^{IJ}$-operator? How should particle creation and annihilation be handled? Etc. Simply put: further study is needed...

Finally, a summary of the summary: The scientific-minded reader may argue that all these constructs are a lot more complicated than simply accepting space-time curvature as a postulate. The point is, however, that if the ideas suggested here are correct, then the various mappings defined in this paper are, in a sense, no longer needed.

## *References*

[1] Dirac, Proc. Roy. Soc. London, Vol 136, pg 453, 1932
[2] A.Einstein "Zur allgemeinen Relativitätstheorie" *Königlich Preußische Akademie der Wissenschaften* (Berlin). *Sitzungsberichte* (1915): 778-786, 1915